\documentclass[apj]{emulateapj}
\usepackage{natbib}
\usepackage[]{amsmath}
\newcommand{\bW}{\ensuremath{{\mathbf W}}}
\newcommand{\bF}{\ensuremath{{\mathbf F}}}
\newcommand{\bFt}{\ensuremath{{\mathbf F}^{\rm T}}}
\newcommand{\vecb}{\ensuremath{\vec{\beta}}}
\newcommand{\vecv}{\ensuremath{\vec{v}}}
\newcommand{\vecu}{\ensuremath{\vec{u}}}
\newcommand{\vect}{\ensuremath{\vec{t}}}

\newcommand{\vecsig}{\ensuremath{\vec{\sigma}}}

\newcommand{\beps}{\ensuremath{\boldsymbol{\epsilon}}}
\newcommand{\sw}{\ensuremath{\sin \omega_*}}
\newcommand{\cw}{\ensuremath{\cos \omega_*}}
\newcommand{\cO}{\ensuremath{\cos \Omega}}
\newcommand{\sO}{\ensuremath{\sin \Omega}}
\newcommand{\ci}{\ensuremath{\cos i}}
\newcommand{\si}{\ensuremath{\sin i}}

\shorttitle{Fitting Keplerian Curves}
\shortauthors{Wright \& Howard}

\begin{document}
\title{Efficient Fitting of Multi-Planet Keplerian Models to Radial
  Velocity and Astrometry Data}
\author{J.~T.~Wright}
\affil{226 Space Sciences Building, Astronomy Department, Cornell University, Ithaca, NY 14853\\email:~jtwright@astro.cornell.edu}
\and
\author{A.~W.~Howard\altaffilmark{1}}
\affil{601 Campbell Hall, Astronomy Department, University of California, Berkeley, CA 94720}

\altaffiltext{1}{Townes Fellow, Space Sciences Laboratory, UC Berkeley}

\begin{abstract}
  We describe a technique for solving for the orbital elements of
  multiple planets from radial velocity (RV) and/or astrometric data taken
  with 1 m/s and $\mu$as precision, appropriate for efforts to
  detect Earth-massed planets in their stars' habitable zones, such as
  NASA's proposed Space Interferometry Mission.  We include details of
  calculating analytic derivatives for use in the Levenberg-Marquardt (LM)
  algorithm for the problems of fitting RV and astrometric data
  separately and jointly.  

  We also explicate the general method of separating the linear and
  nonlinear components of a model fit in the context of an LM fit,
  show how explicit derivatives can be calculated in such a model, and
  demonstrate the speed up and convergence improvements of such a 
  scheme in the case of a five-planet fit to published radial velocity
  data for 55 Cnc. 
\end{abstract}


\section{Introduction}
\subsection{Fitting Kepelerian Curves}
The discovery of over 27 multiple-planet systems in recent years
\citep{Wright09} has required algorithms for
disentangling the radial-velocity signature of such complex systems.
Because the parameters describing a radial velocity (RV) or
astrometric curve are nonlinear, there is no way to fit for them 
analytically, and they must be found through an algorithmic search.
Fitting for the Keplerian parameters of a single orbital companion is usually
straightforward given a good period guess, and if
necessary a ``brute force'' mapping of the $\chi^2$ space is usually not 
computationally prohibitive.  Fitting multiple planets involves 
searching a correspondingly higher-dimensional space and can require
substantial computing time.  

There is an art to searching such  $\chi^2$ spaces efficiently, and in
this context there are many ``tricks'' for finding the global minimum.
For instance, a Lomb-Scargle periodogram \citep{Scargle82} is often used 
to identify promising periods for prospective planets, and all of the
tallest peaks can be used as starting guesses for the fitting
algorithm.  In hierarchical systems, the dominant planet can be fit
for alone, its signal subtracted from the data, and additional planets
can be searched for among the residuals.  This process can then be
repeated until all of the planets have been identified, and then a full,
multi-planet fit on the original data starting at the values found
for the individual planets.\citep[e.g.][]{Fischer08}.  

In many multi-planet systems, planet-planet interactions can
significantly alter the radial velocity (RV) or astrometric signature of the system.  In
such cases where interactions are important, a full dynamical
(Newtonian) fit involving an $n$-body code must be used to properly
fit the data and to ensure the short- and long-term stability of the
solution.  Even in these cases, a multi-planet Keplerian (kinematic)
fit, which simply adds the reflex signatures of single planets and
ignores planet-planet interactions, is still useful for efficiently
identifying planets and providing good initial guesses to the n-body
codes. 

The Levenberg-Marquardt method \citep[LM;][]{Levenberg44,Marquardt63}
is an efficient algorithm for finding a local minimum in a nonlinear
$\chi^2$ space (given good guesses), and is well-suited for the
application of RV and astrometric fitting \citep{Press92}.  As a
concrete example, we refer in this work to a useful IDL\footnote{IDL
  is a commercial programming language and 
  environment by ITT Visual Information Solutions.  http$:$//www.ittvis.com/idl/}
implementation of this technique, \texttt{MPFIT} by Craig
Markwardt\footnote{Available at  http$:$//purl.com/net/mpfit.  \texttt{MPFIT} is a port of \texttt{MINIPACK-1} from FORTRAN, and is also available in C and Python.}
\citep{Markwardt09}.
 Like most implementations of the algorithm, \texttt{MPFIT} requires a
user-defined function that accepts, as an argument, trial values for
the parameters being solved for, evaluates the model at those values,
and returns the corresponding residuals to the data.  \texttt{MPFIT}
then uses this information to step through parameter space and locate
the minimum in $\chi^2$ using a combination of Newton's method and a 
steepest-descent method.  Uncertainties in these parameters can then
be calculated by mapping the $\chi^2$ space near this minimum
\citep[e.g.][]{Wright07} or through error ``bootstrapping''
\citep[e.g.][]{Butler06}. 

The user-defined function in \texttt{MPFIT} also optionally returns the
values of the derivatives of these residuals with respect to the
parameters being fit, computed ``analytically'' or ``explicitly''.
Absent these derivatives, \texttt{MPFIT} will calculate numerical derivatives with small steps in the fitted
parameters and calculating the resulting change in the residuals. he
user must choose step sizes that are not too large --- thereby missing
fine structure in the $\chi^2$ space --- or too small --- increasing
compute time and potentially losing numerical precision.  Adding explicit-derivative capability to
\texttt{MPFIT}, or any implementation of the LM algorithm, obviates the
need for explicitly providing step-sizes, and can greatly improve
performance in terms of the number of steps taken and the CPU time
consumed per step.   

The LM algorithm is useful for finding the best-fitting parameters of
a nonlinear model to a set of data from a ``frequentist'' perspective.
A Bayesean approach can provide for more robust estimates of parameter
uncertainties, especially when those uncertainties are large.  One
Bayesean method of exploring complex or high-dimensional spaces is the
Markov chain Monte Carlo method
\citep[MCMC;][]{Metropolis53,Hastings70}, which has been productively
employed in the context of orbital fitting \citep[e.g.][]{Ford04b,
  Driscoll05, Balan08}. 


Future space missions, such as NASA's Space Interferometry Mission
(SIM Lite), will obtain $\mu$as 
astrometry of nearby stars, referenced to an inertial astrometric
grid.  These measurements will be sufficiently precise to detect
Earth-mass planets with orbital periods shorter than the mission
lifetime.  Since multiple-planet systems are common,\footnote{At
  least 1/4 of known planetary systems show evidence of multiple
  companions \citep{Wright07}} interpreting these data in conjunction
with precise radial velocity data will require algorithms that can
efficiently and robustly search the large nonlinear parameter
space of multiple-planet systems. 

In this paper, we describe the method of efficiently fitting
multi-Keplerian models to such high precision RV and astrometric data
separating the parameters into linear and nonlinear components
in the context of a LM algorithm, and provide the explicit
derivatives used in such a fit.  With modifications, the principles
here can also be applied to MCMC methods, as well
\citep[e.g.]{Catanzarite09, Bakos09}.

\subsection{Plan}
We begin with an elementary exposition to familiarize the reader with
our notation (detailed in Table~\ref{variables}) and provide context
for the later discussion.
In \S\ref{Example}, we explicate the method of exploiting linear parameters
in the Kepler problem using the example of an RV time series.  The
calculation of 
explicit derivatives in \S\ref{explicit} is general to any application
of the LM algorithm where the model contains both linear and nonlinear
parameters.  We apply this method to the problem of astrometric data
in \S\ref{General} and specifically to the problem of combined
astrometric and RV data in \S\ref{combine}.  We discuss nonlinear
terms relevant to $\mu$as astrometric work and how to accommodate them
in \S\ref{nonlinear}.  In \S\ref{improvement} we quantify the
improvement in speed and convergence from exploiting linear parameters and
from the use of explicit derivatives in the LM algorithm.

\section{An Example: Radial Velocities}
\label{Example}
\subsection{Costs and Benefits of Exploiting Linear Parameters}

When fitting RV data, there are $5n+1$ Keplerian parameters to be fit,
where $n$ is the number of planets:\footnote{Occasionally, an
  additional ``trend''  parameter is used to fit out long-term RV trends caused by massive, long-period companions. The other two elements,
  $i$, the inclination, and $\Omega$, the position angle of the ascending
  node, can only be determined astrometrically.}
$P$, the period of the planet's orbit; $K$, the semi-amplitude of the
radial velocity signal; $e$, the eccentricity of the orbit; $\omega$, the
  argument of periastron; $t_{\rm p}$, a date of periastron passage;
  and $\gamma$, the apparent radial velocity of the center of mass of
  the system\footnote{In practice, $\gamma$ is degenerate with an
    arbitrary instrument-dependent RV offset.}

Exploiting linear parameters, as described below, reduces the search space
to $3n$ dimensions (corresponding to $P$, $e$, and $t_{\rm p}$ for 
each planet) and combines the other orbital elements into a set of
linear parameters, which can be solved analytically (and therefore
quickly and exactly) at each step in the search.  The exact, analytic
solution of these linear parameters greatly increases the speed
and stability of search algorithms, but at a cost: the nonlinear
parameters cannot be varied independently of the linear
parameters.  

Further reduction in the number of nonlinear parameters per planet is
certainly possible.  By exploiting an epicyclic or harmonic series expansion
one can reduce the problem to only one nonlinear parameter per planet,
$P_j$.  \citet{Cumming03} and \citet{Ford08b} analyze the RV problem
in the case of a circular orbits, and discuss the relationship between
periodograms and Bayesian approaches to orbit fitting.
\citet{Konacki99} pursued a method analogous to ours in their
approach to RV curve fitting, and \citet{Konacki02} did the same for
astrometry.  Such approximations offer a different set of costs and
benefits to the one presented here.  For instance, when $K$ and $e$
are both large, it may require a large number of terms to adequately
describe a set of RV data.  We may pursue such an approach in a future
version of our code. 

The linear parameters in our treatment are not coefficients in a
series expansion; rather, we recast the problem, separating
linear parameters, which can be solved for exactly with linear
algebra, from nonlinear parameters, which must be 
solved for algorithmically (using LM).  As described in \S2.2, the
linear parameters are algebraic combinations of $K$, $\omega$,
$\gamma$, and an optional trend parameter, while $P$,
$t_{\mathrm{P}}$, and $e$ are nonlinear parameters.

Two complications are introduced by exploiting linear parameters.
First, because the linear parameters are computed analytically, and
not algorithmically, their errors and covariances with the nonlinear
parameters are not computed automatically in the procedure outlined
here.  (They also cannot be held 
fixed in the fit, although the issue of fixing the trend parameter can
be finessed -- see \S2.6).  The second complication is that computation
of the explicit derivatives for the LM algorithm is not
straightforward: in the context of the algorithm, the Keplerian model
is not a simple function of the 3$n$ nonlinear parameters being fit,
but also depends on the data.  For example, increasing the nonlinear
parameter $e$ slightly not only changes the model because it is more
eccentric, but also because at this new value of $e$ the linear
parameters have different values.  These complications can be avoided
if one uses the linear basis of the orbits presented here in an
ordinary, nonlinear 5$n$ + 1 parameter fit, but the convergence and
speed-up benefits will not be as great. 

The sections below describe a general method of calculating an
$n$-planet Keplerian radial velocity model given values for the $3n$
nonlinear parameters, and of calculating the derivatives of that model
with respect to those parameters.  These equations can be used in a
user-defined function for input to an LM minimization code, or a
brute-force $\chi^2$ map. 

\subsection{Linear Parameters in the Radial Velocity Problem}
\label{RV}
We wish to find the parameters of the n-planet model
\begin{equation}
\label{modeleq}
u(t) = \sum_{j=1}^n [K_j(\cos(\omega_j+f_j(t))+e_j\cos\omega_j)] +
\gamma + d \cdot (t-t_0)
\end{equation}
which best fits the set of observed radial velocities \vecv, measured at times
\vect, with uncertainties \vecsig\ in a least-squares sense.
Here,
$K_j$ and $\omega_j$ are the usual Keplerian parameters for planet $j$,
$\gamma$ is the time-independent velocity offset, 
$d$ is the trend parameter, $f_j(t)$ is the true anomaly of planet
$j$ at time $t$, and $t_0$ is a conveniently chosen epoch of the
observations.  The true anomaly is defined implicitly in terms of 
the other three Keplerian parameters ($P_j,t_{{\rm p},j}$, and $e_j$) through
the relations  
\begin{equation}
\label{edef}
\tan\frac{f_j(t)}{2} = \sqrt{\frac{1+e_j}{1-e_j}} \tan \frac{E_j(t)}{2}
\end{equation}
\begin{equation}
\label{Kepler}
E_j(t)-e_j \sin E_j(t) = \frac{2\pi(t-t_{{\rm p},j})}{P_j} = M_j(t)
\end{equation}
In Eq.~\ref{Kepler} (known as Kepler's Equation) $E_j$ is called the
{\it eccentric anomaly} of planet $j$, and $M$ is known
as the {\it mean anomaly}.\footnote{Instead of $t_{\rm p}$, many
  authors (especially dynamicists) prefer to parameterize orbits in
  terms of the {\it mean longitude at epoch}, defined as $M(t_0)+\Omega+\omega$.} 

We identify the linear parameters by rewriting Eq.~\ref{modeleq} as 
\begin{equation}
\label{linmodel}
  u(t) = \sum_{j=1}^n [h_j \cos f_j(t) + c_j \sin f_j(t)] + v_0 + d
  \cdot (t-t_0)
\end{equation}
where
\begin{equation}
\label{adef}
h_j = \phantom{-}K_j \cos \omega_j
\end{equation}
\begin{equation}
\label{bdef}
c_j = -K_j \sin \omega_j
\end{equation}
and
\begin{equation}
\label{cdef}
v_0 = \gamma + \sum_{j=1}^n K_j e_j \cos\omega_j
\end{equation}

These linear parameters can be converted back to Keplerian orbital
elements through the relations
\begin{equation}
\label{defk}
K_j= \sqrt{h_j^2+c_j^2}
\end{equation}
\begin{equation}
\tan \omega_j = \frac{-c_j}{\phantom{-}h_j}
\end{equation}
(where $\omega_j$ is chosen so that $\sin \omega_j$ has the sign of
the numerator) and 
\begin{equation}
\label{defgam}
\gamma = v_0 - \sum_{j=1}^n K_j e_j \cos\omega_j
\end{equation}
The masses of the orbiting planets can be inferred from their corresponding
semi-amplitudes $K_j$, defined for a single-planet system with planet mass
$m$ and stellar mass $m_*$ as
\begin{equation}
\label{defmf}
K^3= \frac{2\pi G}{P(1-e^2)^{\frac{3}{2}}}\left(\frac{m^3\sin^3i}{(m_*+m)^2}\right)
\end{equation}
where $G$ is Newton's gravitational constant.  The fraction in
parentheses is known as the {\it mass function} of the 
system.

The problem can now be divided into two parts:  an
algorithmic search through parameter space for the best-fit nonlinear
parameters $P_j$, $e_j$, and $t_{{\rm p},j}$ with a computer routine
such as an LM or an MCMC code, and at each step in that search an
analytic solution for the linear parameters that fit best there.

\subsection{Solving for the Linear Parameters}
Given a set of values for the nonlinear parameters, we can
fit for the linear parameters in Eq.~\ref{linmodel} through $\chi^2$
minimization.  We denote the row vector of linear parameters
\begin{equation}
\label{defbeta}
\vecb = \{h_1,c_1,h_2,c_2 \ldots h_n,c_n,v_0,d\}
\end{equation}
We define $\chi^2$ the usual way: 
\begin{equation}
\chi^2 = \sum_{k=1}^{N} \frac{(v_k - u(t_k))^2}{\sigma_k^2} 
\end{equation}
and minimize it with respect to each of the linear parameters in
\vecb\ simultaneously:
\begin{equation}
\label{chi2}
\frac{\partial\chi^2}{\partial \beta_l} = -2\sum_{k=1}^{N} \frac{v_k - u(t_k)}{\sigma_k^2}
\left.\frac{\partial\, u\,}{\partial \beta_l\!\!}\,\right|_{t=t_k} \!\!\!\! = 0 
\end{equation}
We can express this more compactly by invoking matrix algebra. 
For the problem of Keplerian orbits, we define the matrix \bF\ as
\begin{equation}
\label{deff}
\bF =
\left[ 
  \begin{array}{ccc} 
    \cos f_{1,1} & \cos f_{1,2} & \ldots \\
    \sin f_{1,1} & \sin f_{1,2} & \ldots \\
    \cos f_{2,1} & \cos f_{2,2} & \ldots \\ 
    \sin f_{2,1} & \sin f_{2,2} & \ldots \\ 
    \vdots & \vdots &\\
    \cos f_{n,1} & \cos f_{n,2} & \ldots \\ 
    \sin f_{n,1} & \sin f_{n,2} & \ldots \\ 
     1  &  1  & \ldots\\
    t_1-t_0 & t_2-t_0 & \ldots\\
  \end{array}
 \right]
\end{equation}
where $f_{j,k} \equiv f_j(t_k)$.  This allows us to write the model velocities at times
\vect\ (Eq.~\ref{modeleq}) as  
\begin{equation}
\label{modeleqmat}
\vecu=\vecb\bF
\end{equation}
We also define the diagonal weight matrix \bW\ such that 
\begin{equation}
\label{defKron}
W_{kl} = \delta_{kl}/\sigma_k^2
\end{equation}
where we have used the Kronecker delta symbol.  We can then write the
system of equations in Eq.~\ref{chi2} as
\begin{equation}
\frac{\partial\chi^2}{\partial \vecb} = -2(\vec{v}-\vecb \bF) \bW \bFt = \vec{0}
\end{equation}
Solving for \vecb\ we have
\begin{equation}
\label{lssol}
\vecb = \vec{v}\bW\bFt\beps
\end{equation}
where we have denoted the error matrix (also called the
variance-covariance matrix)
\begin{equation}
\label{beps}
\beps = (\bF \bW \bFt)^{-1}
\end{equation}

Eq.~\ref{lssol} represents the general solution to the linear
least-squares problem for an appropriately defined \bF.  In the
context of Keplerian fits, given a set of nonlinear parameters $P$,
$e$, and $t_{\rm p}$ for each planet, the remaining Keplerian orbital
elements can be found from \vecb\ using
Eqs.~\ref{defk}--\ref{defgam}. 

\subsection{Calculation of Explicit (Analytic) Derivatives for Use in the LM Algorithm}
\label{explicit}
The derivative of the model velocities \vecu\ with respect to any
nonlinear parameter can be found from Eq.~\ref{modeleqmat}:
\begin{equation}
\label{dudx}
\frac{d \vecu}{d x} = \frac{d \vecb}{d x}\bF+
\vecb\frac{d \bF}{d x}
\end{equation}
where $x$ stands for any of the nonlinear parameters (here, $P_j$,
$e_j$, or $t_{{\rm p},j}$).  

From Eq.~\ref{lssol} we have
\begin{equation}
\label{dadx}
\frac{d \vecb}{d x} = \vecv\bW\left(\frac{d \bF}{d
  x}^{\rm T}\beps+\bFt\frac{d \beps}{d x}\right)
\end{equation}
From the definition of a matrix inverse we can express the last derivative as
\begin{equation}
\frac{d \beps}{d x} = -\beps\left(\frac{d}{d x}\beps^{-1}\right)\beps
\end{equation}
Using Eq.~\ref{beps} we then have
\begin{eqnarray}
\label{endlin}
\frac{d \beps}{d x} &=& -\beps\frac{d}{d x}\left(\bF\bW\bFt\right)\beps\\
&=&-\beps\left(\left(\frac{d \bF}{d x}\bW\bFt\right)+\left(\frac{d \bF}{d x}\bW\bFt\right)^{\rm T}\right)\beps
\end{eqnarray}
Eqs.~\ref{modeleqmat}--\ref{endlin} are not specific to the RV Kepler
problem, but are a general method of calculating explicit
derivatives in a model with both linear and nonlinear parameters, and so
can be applied to any analogous problem.  For instance, the problem of
fitting an orbit using astrometric data also has linear parameters as
we show in \S\ref{General}.  

\subsection{Explicit Derivatives for the Radial Velocity Model}
For the case of a Keplerian RV model, from Eq.~\ref{deff} we have
\begin{equation}
\frac{d\bF}{d x} =
\left[ 
  \begin{array}{ccc} 
             - \sin f_{1,1}f^{\prime}_{1,1}&          - \sin f_{1,2}f^{\prime}_{1,2}&\ldots \\
    \phantom{-}\cos f_{1,1}f^{\prime}_{1,1}& \phantom{-}\cos f_{1,2}f^{\prime}_{1,2}&\ldots \\
             - \sin f_{2,1}f^{\prime}_{2,1}&          - \sin f_{2,2}f^{\prime}_{2,2}&\ldots \\
    \phantom{-}\cos f_{2,1}f^{\prime}_{2,1}& \phantom{-}\cos f_{2,2}f^{\prime}_{2,2}&\ldots \\
    \vdots & \vdots& \\
             - \sin f_{n,1}f^{\prime}_{n,1}&          - \sin f_{n,2}f^{\prime}_{n,2}&\ldots \\
    \phantom{-}\cos f_{n,1}f^{\prime}_{n,1}& \phantom{-}\cos f_{n,2}f^{\prime}_{n,2}&\ldots \\
     0 & 0 & \ldots \\
     0 & 0 & \ldots 
  \end{array}
 \right]
\end{equation}
where $f^{\prime}_{j,k} \equiv df_{j,k} / d x$.  Note that since $f^{\prime}_{j,k}$ refers to
the true anomaly of planet $j$, it vanishes when $x$ refers to a
parameter of a different planet (e.g., $d f_{j,k} / d P_l = 0$ when $j
\neq l$).  This means that the matrix $d\bF/d x$ has only two nonzero rows.  We
can therefore suppress subscripts below for clarity. 

We can calculate the nonzero derivatives as
\begin{equation}
  \label{dfdx}
  \frac{d f}{d x} = \frac{\partial f}{\partial x}+\frac{\partial f}{\partial E}\frac{d E}{d x}
\end{equation}
where, from Kepler's Equation (Eq.~\ref{Kepler}), we have
\begin{equation}
\label{dedp}
  \frac{d E}{d P} = \frac{-2\pi(t-t_{\rm p})/P^2}{1-e\cos E}
\end{equation}
\begin{equation}
  \frac{d E}{d t_{\rm p}} = \frac{-2\pi/P}{1-e\cos E}
\end{equation}
\begin{equation}
  \label{dede}
  \frac{d E}{d e} = \frac{\sin E}{1-e\cos E}
\end{equation}
and from Eq.~\ref{edef} we have
\begin{equation}
\frac{\partial f}{\partial E} = \sqrt{\frac{1+e}{1-e}}\frac{1+\cos f}{1+\cos E}
\end{equation}
From Eq.~\ref{edef} we can also write
\begin{equation}
  \frac{\partial f}{\partial e}=\frac{2\tan(E/2)\cos^2(f/2)}{(1-e)\sqrt{1-e^2}}
\end{equation}
but it is more computationally convenient to note that this happens to
simplifiy to
\begin{equation}
  \frac{\partial f}{\partial e}=\frac{\partial f}{\partial E}\frac{\sin
  E}{1-e^2}
\end{equation}
Finally,
\begin{equation}
\label{last}
  \frac{\partial f}{\partial P}=\frac{\partial f}{\partial t_{\rm p}}=0
\end{equation}

Eqs.~\ref{dadx}--\ref{last} can be used to calculate the terms in
Eq.~\ref{dudx}, yielding the explicit derivatives used by LM method
fitting routines, such as \texttt{MPFIT}.

\subsection{Variations on the $n$-Planet RV Model}
\label{variations}
In the context of the Kepler problem, the above equations include a
parameter for a linear trend in the data.  In practice, fitting for
such a trend will only occasionally be necessary.  When not needed the
$d$ parameter in \vecb\ and the bottom rows of the \bF\ and $d\bF/d x$
matrices can simply be left out of the calculations.   Similarly, the
trend parameter can effectively be held fixed at a nonzero value by
simply subtracting the desired value from the data before fitting.  

These matrices can also be easily extended to handle the case of combining
data from multiple telescopes between which there exist RV offsets.
This is accomplished by extending the data vectors \vecv, \vect, and
\vecsig\ to include data from all telescopes, and extending the vector
\vecb\ to include a separate offset parameter for each telescope after
the first.  The corresponding rows of \bF\ must then be filled with
1's in those columns corresponding to data from the appropriate
telescope, and 0's elsewhere.  Naturally, in $d\bF / dx$ the elements
of these rows are all 0. 

\section{Applications to Astrometry}
\label{General}

\subsection{Astrometry Alone}
\label{astrometryalone}
The above method can also be extended to the problem of fitting for
the Keplerian elements of an orbit from astrometric data.  Here we
present a simplified model of $\mu$as astrometric data of the sort
that might be provided by a space observatory such as NASA's Space
Interferometry Mission (SIM Lite).  We anticipate modifying our algorithms
for a more realistic model of interferometric data and its noise
sources \citep{Sozzetti05,Eriksson07,Catanzarite08} in the near future.

The linear basis for astrometric coordinates are the Thiele-Innes
constants \citep[e.g.][]{Binnendijk60}, and are well documented as
useful tools for astrometric curve-fitting \citep[e.g.][]{Casertano08}. The
astrometric perturbations caused by an orbiting companion can be
described in terms of six astrometric orbital elements: in addition to
$e$ and $t_{\rm p}$, we have $a$, the semi-major axis of the star's
apparent orbit on the sky in units of arc; $\Omega$, the longitude of
the ascending (approaching) node (measured as a position angle on the
sky); $i$, the inclination of the orbit on the sky (such that $i=0$
corresponds to a face-on, clockwise orbit); and $\omega_*$, the longitude of periastron of the {\it
star's} orbit.\footnote{The orbital parameters of the star and those
of the unseen companion are all identical except $a$, which differs
by a factor of $m/m_*$, and $\omega$, which differs by $\pi$.}

The Thiele-Innes constants are defined in
terms four of the astrometric elements of the star's orbit about the
secondary: 

\begin{eqnarray}
A &=& a(\phantom{-}\cO \cw - \sO\sw\ci) \\ 
B &=& a(\phantom{-}\sO \cw + \cO\sw\ci)\\
F &=& a(-\cO \sw - \sO\cw\ci)\\
G &=& a(-\sO \sw + \cO\cw\ci)\\
C &=& a\sw \si \\
H &=& a\cw \si
\end{eqnarray}
These constants can be quickly computed using rotation matrices:
\begin{equation}
\label{defTI}
\left[ 
  \begin{array}{ccc}
A & B & C\\
F & G & H\\
a\sin i\sin \Omega & -a\sin i \cos \Omega & a\cos i
  \end{array}
 \right] = a R_z(\omega_*) R_x(i) R_z(\Omega)
\end{equation}
where $R$ is the 3-D rotation matrix
\begin{eqnarray}
\label{defr}
R_z(\Omega) = \left[ 
 \begin{array}{ccc}
   \phantom{-}\cos \Omega & \phantom{-} \sin \Omega & \phantom{in}0\phantom{-} \\ 
   -\sin \Omega & \phantom{-}\cos \Omega & \phantom{in}0\phantom{-} \\
   0 & 0 & \phantom{in}1\phantom{-}
   \end{array}
\right]
\end{eqnarray}
and  $\omega_*$ is the argument of periastron of the orbit of the {\it
  star}.  

We can transform the Thiele-Innes
constants back to Keplerian orbital elements of the planet with the relations:
\begin{eqnarray}
\label{sumom}
\tan (\omega_*+\Omega) &=& \phantom{-(}\frac{B-F}{A+G}\\
\label{difom}
\tan (\omega_*-\Omega) &=& \frac{-(B+F)}{\phantom{-(}A-G\phantom{)}}\\
\tan^2\left(\frac{i}{2}\right) &=&
\frac{(A-G)\cos(\omega_*+\Omega)}{(A+G)\cos(\omega_*-\Omega)}\
\end{eqnarray}
\begin{equation}
\begin{array}{rl}
a =&(A\cos\omega_*-F\sin\omega_*)\cos\Omega-\\
&(A\sin\omega_*+F\cos\omega_*)\sin\Omega\sec i
\end{array}
\end{equation}
and
\begin{equation}
\omega=\omega_*+\pi
\end{equation}
where the quadrants of $\omega_*-\Omega$ and $\omega_*+\Omega$ are
determined by the signs of the numerators in Eqs.~\ref{sumom} \&
\ref{difom}.  These relations leave a $\pm \pi$ ambiguity in
$\omega_*$, $\omega$, and $\Omega$ that can only be resolved by radial velocities, without
which convention dictates that we choose the solution with $\Omega < \pi$. 

The Thiele-Innes constants $C$ and $H$ are closely related
to the $c$ and $h$ constants of Eqs.~\ref{adef} \& \ref{bdef}.  The
set of six constants have the identity 
\begin{equation}
a^2 = A^2+B^2+C^2=F^2+G^2+H^2
\end{equation}

In astrometry there are five parameters that describe a star's motion in the
absence of orbiting companions: $\Delta \alpha_0 \cos
\delta$ and $\Delta \delta_0$, the difference between the true and
nominal position of the system at $t_0$; $\mu_\alpha$ and $\mu_\delta$, the proper motions
in the RA and Dec directions; and $\varpi$, the parallax of the
system.  Our model for the astrometric displacement of a star due to
parallax, proper motion, and a system of unseen planets in terms of
the Thiele-Innes constants at times $\vec{\tau}$ is:
\begin{equation}
\label{defda}
\begin{array}{rcl}
\Delta \delta_k &=& \sum_{j=1}^n [A_jX_{j,k}+F_jY_{j,k}] +\\
&& \Delta \delta_0 + \varpi\Pi_{\delta,k} + \mu_\delta(\tau_k-t_0)
\end{array}
\end{equation}
\begin{equation}
\begin{array}{rcl}
\label{defdd}
\Delta \alpha_k \cos \delta &=& \sum_{j=1}^n[B_jX_{j,k}+G_jY_{j,k}]+\\
&&\Delta \alpha_0 \cos \delta +
\varpi\Pi_{\alpha,k} +\mu_\alpha(\tau_k-t_0)
\end{array}
\end{equation}
where $X$ and $Y$ are the so-called elliptical rectangular
coordinates, defined as 
\begin{eqnarray}
\label{defx}
X_{j,k} & = & \cos E_j(\tau_k) - e_j\\
\label{defy}
Y_{j,k} & = & \sqrt{1-e_j^2} \sin E_j(\tau_k)
\end{eqnarray}
where $E$ is the eccentric anomaly and the quantities $\Pi_{\alpha,k}$
and $\Pi_{\delta,k}$ refer to the astrometric displacements due to
parallax in the $\alpha$ and $\delta$ directions\footnote{In
  this paper the bare symbols $\alpha$ and $\delta$ will always refer
  to the nominal right ascension and declination of a star at $t_0$, absent the
  effects of parallax and astrometric displacement from companions.}
 during observation $k$, which are given by [cite Supplement to
   Astronomical Almanac here]:  
\begin{equation}
\label{defpi}
\Pi_{\alpha,k} = r_x(\tau_k)\sin\alpha-r_y(\tau_k)\cos\alpha
\end{equation}
\begin{equation}
\label{defpi2}
\Pi_{\delta,k} = \left(r_x(\tau_k)\cos\alpha+r_y(\tau_k)\sin\alpha\right)\sin\delta-r_z(\tau_k)\cos\delta
\end{equation}
Here $(r_x,r_y,r_z)$ represent the Cartesian components in equatorial
coordinates of the position of the observatory, $\vec{r}$, at time
$\tau$ with respect to the Solar System barycenter  (in units of 
AU when $\varpi$ is in arcsec).  These values for the Earth are
available from the NASA Jet Propulsion Laboratory Solar System
ephimerides,\footnote{http$://$ssd.jpl.nasa.gov} but for $\mu$as 
and spaceborne work the precise position of the observatory itself is
required. 
  
Note that since $a$ is the apparent semi-major axis of the star's orbit
in units of arc, its relationship to the mass of the secondary depends on
the method of astrometry used.  For astrometric perturbations due to
an unseen planet, (that is, absolute astrometric displacements with
respect to the sidereal frame, as measured by SIM Lite) we have from Kepler's Third Law
\begin{equation}
\label{defastar}
a^3 = \frac{\varpi^3m^3}{(m_*+m)^2}P^2
\end{equation}
where $a$ is measured in arcseconds when $P$ is measured in years,
$\varpi$ is the parallax in arcseconds, and $m$ is the mass of the unseen 
companion and $m_*$ is the mass of the star in solar massses.\footnote{\label{relastro}In the
case where the binary orbit is measured as a separation and position
angle of one star with respect to another (i.e., relative astrometry)
the measured separation is given by $a^3 = \varpi^3(m_*+m)P^2$.  The
application of the techniques here for multiple-planet systems 
with relative astrometry is not straightforward, and is beyond the scope
of this manuscript.}

We can now extend Eq.\ref{modeleqmat} to the case of 2-D data by
defining our vector of $N$ measurements taken at times $\vec{\tau}$:
\begin{equation}
\label{defastrou}
[\Delta \delta_1, \Delta \delta_2 \ldots \Delta \delta_N, \Delta \alpha_1
    \cos \delta, \Delta \alpha_2
    \cos \delta \ldots \Delta \alpha_N \cos \delta ]
\end{equation}
and our model with linear parameters for $n$ planets:
\begin{equation}
\label{defastrobeta}
  \begin{array}{cll}
  \vecb &=[A_1,B_1,F_1,G_1\ldots A_n,B_n,F_n,G_n,&  \\
    &\Delta \delta_0,\Delta \alpha_0 \cos\delta,\mu_\delta,\mu_\alpha,\varpi]&
  \end{array}
\end{equation}
and matrix $\bF =$
 \begin{equation}
   \label{defastrof}
\left[
    \begin{array}{cccccccc}
  X_{1,1} & X_{1,2} & \ldots & X_{1,N} & 0      & 0       & \ldots & 0     \\  
  0      & 0       & \ldots & 0      & X_{1,1} & X_{1,2} & \ldots & X_{1,N}\\  
  Y_{1,1} & Y_{1,2} & \ldots & Y_{1,N} & 0      & 0       & \ldots & 0     \\  
  0      & 0       & \ldots & 0      & Y_{1,1} & Y_{1,2} & \ldots & Y_{1,N}\\  
  X_{2,1} & X_{2,2} & \ldots & X_{2,N} & 0      & 0       & \ldots & 0     \\  
  0      & 0       & \ldots & 0      & X_{2,1} & X_{2,2} & \ldots & X_{2,N}\\  
  Y_{2,1} & Y_{2,2} & \ldots & Y_{2,N} & 0      & 0       & \ldots & 0     \\  
  0      & 0       & \ldots & 0      & Y_{2,1} & Y_{2,2} & \ldots & Y_{2,N}\\  
  \vdots & \vdots & & \vdots & \vdots & \vdots& & \vdots\\
  X_{n,1} & X_{n,2} & \ldots & X_{n,N} & 0      & 0       & \ldots & 0     \\  
  0      & 0       & \ldots & 0      & X_{n,1} & X_{n,2} & \ldots & X_{n,N}\\  
  Y_{n,1} & Y_{n,2} & \ldots & Y_{n,N} & 0      & 0       & \ldots & 0     \\  
  0      & 0       & \ldots & 0      & Y_{n,1} & Y_{n,2} & \ldots & Y_{n,N}\\
  1 & 1 & \ldots & 1 & 0 & 0 & \ldots & 0\\
  0 & 0 & \ldots & 0 & 1 & 1 & \ldots & 1\\
  \tau_1-t_0 & \tau_2-t_0 & \ldots & \tau_N-t_0 & 0 & 0 & \ldots & 0\\
  0 & 0 & \ldots & 0 & \tau_1-t_0 & \tau_2-t_0 & \ldots & \tau_N-t_0\\
  \Pi_{\delta,1} & \Pi_{\delta,2} & \ldots & \Pi_{\delta,N} & \Pi_{\alpha,1} & \Pi_{\alpha,2} & \ldots & \Pi_{\alpha,N}
 \end{array}
 \right]
 \end{equation}
The nonzero components of $d\bF/dx$ can be calculated from:
\begin{eqnarray}
\label{dxdp}
  \frac{d X}{d P} &=& - \frac{d E}{d P} \sin E\\
  \frac{d X}{d t_{\rm p}} &=& -\frac{d E}{d t_{\rm p}}\sin E \\  
  \frac{d X}{d e} &=& - \frac{d E}{d e} \sin
  E - 1\\
  \frac{d Y}{d P} &=& \sqrt{1-e^2}\cos E \frac{d E}{d P}\\
  \frac{d Y}{d t_{\rm p}} &=& \sqrt{1-e^2}\cos E \frac{d E}{d t_{\rm p}}\\
\label{dyde}
  \frac{d Y}{d e} &=& \sqrt{1-e^2}\cos E \frac{d
    E}{d e} - \frac{e\sin E}{\sqrt{1-e^2}}
\end{eqnarray}
and Eqs.~\ref{dedp}--\ref{dede}.

\subsection{Astrometry in Arbitrary Coordinates}
Astrometry does not always deliver contemporaneous
($\Delta \alpha \cos \delta$, $\Delta \delta$) pairs at a common time
$t$.  In the general case, a baseline determines the 1-D displacement
of a star from some reference at an intermediate position angle on the
sky (i.e., not necessarily 0 [as in the case for $\Delta \delta$] or
$\pi/2$ [as for $\Delta \alpha \cos\delta$]).  We can combine
Eqs.~\ref{defda} \& \ref{defdd} to a more general form to accommodate
a heterogeneous set of such data :
\begin{equation}
\label{defrho}
\begin{array}{rl}
\rho_{\theta,k} =&\sum_{j=1}^n [(A_j X_{j,k} +F_j
  Y_{j,k})\cos\theta_k+  \\&\phantom{\sum_{j=1}^n [} (B_j X_{j,k}+G_j Y_{j_k})\sin\theta_k] + \\
&(\Delta \delta_0 \phantom{,\cos \delta}+ \varpi\Pi_{\delta,k} + \mu_\delta(\tau_k-t_0)) \cos \theta_k + \\
&(\Delta \alpha_0 \cos \delta + \varpi\Pi_{\alpha,k} +\mu_\alpha(\tau_k-t_0)) \sin \theta_k
\end{array}
\end{equation}
where $\rho$ is the separation and $\theta$ the position angle of the
measurement.

Interestingly, we could also use the other two Thiele-Innes constants
to achieve the same result by defining a new linear parameter scheme where the astrometric displacements are described as:
\begin{equation}
\begin{array}{rl}
\rho_{\theta,k} =& \sum_{j=1}^n [H_j S_{j,k} +C_j T_{j,k}] + \\
&(\Delta \delta_0 \phantom{,\cos \delta}+ \varpi\Pi_{\delta,k} + \mu_\delta(\tau_k-t_0)) \cos \theta_k + \\
&(\Delta \alpha_0 \cos \delta + \varpi\Pi_{\alpha,k} +\mu_\alpha(\tau_k-t_0)) \sin \theta_k
\end{array}
\end{equation}
where S and T are defined for the $k$th measurement and $j$th planet as
\begin{equation}
  \label{defst}
\begin{array}{l}
  \left[
    \begin{array}{c}
      S_{j,k} \\
      T_{j,k}
    \end{array}
    \right] = \\ \\ \left[
    \begin{array}{cc}
      \phantom{-}\cos(\Omega_j-\theta_k)\csc i_j & -\sin(\Omega_j-\theta_k)\cot i_j \\
      -\sin(\Omega_j-\theta_k)\cot i_j & -\cos(\Omega_j-\theta_k)\csc i_j
    \end{array}
    \right] 
  \left[
    \begin{array}{cc}
      X_{j,k} \\
      Y_{j,k}
    \end{array}
    \right]
\end{array}
\end{equation}
This scheme uses only two linear parameters, instead of
four, per planet, and so will not be as efficient, but it is still
useful because it can be easily combined with the radial velocity
scheme of \S\ref{RV}.  We demonstrate this in \S\ref{combine} with a
procedure that can accommodate any combination of radial velocity and
astrometric data.  

We can recover the parameters $a$ and $\omega$ from the linear parameters $C$
and $H$ given the nonlinear parameter $i$ by: 
\begin{eqnarray}
  a_j^2 &=& \frac{C_j^2+H_j^2}{\sin^2 i_j}\\
  \tan \omega_{j} &=& \frac{-C_j}{-H_j}
\end{eqnarray}
where $\omega_j$ is chosen so $\sin \omega_j$ has the same sign as $-C_j$.

\subsection{Combining Astrometry with Radial Velocities}
\label{combine}
Combining astrometric data with radial velocity data is not as
simple as combining the RV-only and astrometry-only schemes
outlined above, because the six linear parameters ($A, B,
F, G$, and $c$ and $h$) are a combination of
only five Keplerian elements ($K, a, \Omega, \omega$, and $i$), and
the problem is thus overconstrained.  One solution would be to
minimize $\chi^2$ subject to the appropriate constraints using
Lagrange multipliers, but the resulting set of nonlinear
equations may not be guaranteed to have a unique solution and would be
difficult to solve for the case of an arbitrary number of planets.

Another solution is to adapt the $C$ and $H$ constants to accommodate
general astrometric data, and these constants are closely related to
the scheme for RV data used in \S\ref{RV}.  To do this,
we define our vector of measurements to be 
\begin{equation}
[\vecv,\vec{\rho}_{\vec{\theta}}]
\end{equation}
where the velocities are taken at times \vect\ and the astrometry at
times $\vec{\tau}$ , and define our model, as in Eq.~\ref{modeleqmat}:
\begin{equation}
\begin{array}{rl}
\vecb&=[H_1,C_1,H_2,C_2\ldots H_n,C_n,v_0,d,\\
&\Delta \delta_0,\Delta \alpha_0
   \cos\delta,\mu_\delta,\mu_\alpha,\varpi]
\end{array}
\end{equation}
From Eqs.~\ref{adef}, \ref{bdef}, \ref{defTI}, \& \ref{defastar} we have:  
\begin{eqnarray}
\label{defHC}
h_j = -\lambda_jH_j\\
\label{defHC2}
c_j = \phantom{-}\lambda_jC_j
\end{eqnarray}
where we have introduced a
$\lambda$, a combination of nonlinear orbital parameters of planet $j$ which has
units of velocity and has the value
\begin{eqnarray}
\lambda_j=\frac{2 \pi \mbox{AU}}{\tilde{\varpi} P_j \sqrt{1-e_j^2}}
\end{eqnarray}
when the estimated parallax, $\tilde{\varpi}$, is expressed in arcseconds.

The parameters $H_j$, $C_j$, and $v_0$ can be transformed into $m_j^3/(m_*+m_j)^2$, $\omega_j$, and $\gamma$ with
Eqs.~\ref{defk}--\ref{defmf} \& \ref{defHC}--\ref{defHC2}.  

The appearance of $\varpi$ in the definition of $\lambda_j$ (and
therefore in $C_j$ and $H_j$) indicates a  
fundamental nonlinearity in the combined RV-astrometry problem: the
parallax is not a truly linear parameter.  However, because it is
nearly linear, if a good estimate of the parallax is available then
this can be used in the formula for $\lambda_j$ (we indicate the
approximate nature of the parallax term with a tilde).  Once 
the parallax is solved for more precisely, the fit can be re-run with
an improved estimate of $\varpi$.  This procedure should converge very
quickly, and we will use it again to deal with other, smaller nonlinear
terms in \S\ref{nonlinear}. 

The first columns of the \bF\ matrix, corresponding to the radial
velocity measurements, now read 
\begin{equation}
\label{bfrvastro}
\left[ 
\begin{array}{ccc}
 -\cos f_{1,1} \lambda_1 & -\cos f_{1,2} \lambda_1 & \ldots \\
 \phantom{-}\sin f_{1,1} \lambda_1 & \phantom{-}\sin f_{1,2} \lambda_1  & \ldots\\
 -\cos f_{2,1} \lambda_2 & -\cos f_{2,2} \lambda_2 & \ldots \\
 \phantom{-}\sin f_{2,1} \lambda_2 & \phantom{-}\sin f_{2,2} \lambda_2  & \ldots\\
 \vdots              &   \vdots &\\
 -\cos f_{n,1} \lambda_n & -\cos f_{n,2} \lambda_n & \ldots \\
 \phantom{-}\sin f_{n,1} \lambda_n & \phantom{-}\sin f_{n,2} \lambda_n  & \ldots\\
  1 &  1 & \ldots\\
  t_1-t_0 &  t_2-t_0 & \ldots \\
  0 & 0 & \ldots \\
  0  & 0 & \ldots\\
  0  & 0 & \ldots \\
  0  & 0 & \ldots \\
  0  & 0 & \ldots 
\end{array}
\right.
\end{equation}
and the rest of the columns, corresponding to the astrometric
measurements, read
\begin{equation}
\label{bfrvastro2}
\left.
\begin{array}{ccc}
 S_{1,1} &  S_{1,2} & \ldots \\
 T_{1,1} &  T_{1,2} & \ldots \\
 S_{2,1} &  S_{2,2} & \ldots \\
 T_{2,1}&   T_{2,2} & \ldots \\
          \vdots &    \vdots & \\
 S_{n,1} &  S_{n,2} & \ldots \\
 T_{n,1}&   T_{n,2} & \ldots \\
 0 & 0 & \ldots\\
 0 & 0 & \ldots \\
 \cos \theta_1 & \cos \theta_2 & \ldots\\
 \sin \theta_1 & \sin \theta_2 & \ldots\\
 (\tau_1-t_0)\cos \theta_1 & (\tau_2-t_0)\cos \theta_2 & \ldots\\
 (\tau_1-t_0)\sin \theta_1 & (\tau_2-t_0)\sin \theta_2 & \ldots\\
  \Pi_{\delta,1}\cos \theta_1 + \Pi_{\alpha,1}\sin\theta_1 &
  \Pi_{\delta,2}\cos \theta_2 + \Pi_{\alpha,2}\sin\theta_2 & \ldots
\end{array}
\right]
\end{equation}

For the nonlinear parameters $P_j$, $T_{{\rm p},j}$, and $e_j$, the two
nonzero rows of $d\bF/dx$ can be calculated from:

\begin{eqnarray}
\frac{d}{d x}(\phantom{-}\cos f(t) \lambda) = -\sin f(t)
f^\prime(t) \lambda + \cos f(t) \frac{d \lambda}{d x}\\
\frac{d}{d x}(-\sin f(t) \lambda) = -\cos f(t)
f^\prime(t) \lambda - \sin f(t) \frac{d \lambda}{d x}
\end{eqnarray}
where
\begin{eqnarray}
\frac{d \lambda}{d P} &=& -\frac{\lambda}{P}\\
\label{lastlambda}
\frac{d \lambda}{d e} &=&\frac{e\lambda}{1-e^2}\\
\frac{d \lambda}{d \Omega} = \frac{d \lambda}{d i} &=& \frac{d \lambda}{d
  t_{\rm p}} =0
\end{eqnarray}
and
\begin{equation}
\begin{array}{l}
  \left[
    \begin{array}{c}
      d S/d x \\
      d T/d x
    \end{array}
    \right] = \\ \\ \left[
    \begin{array}{cc}
      \phantom{-}\cos(\Omega-\theta)\csc i & -\sin(\Omega-\theta)\cot i \\
      -\sin(\Omega-\theta)\cot i & -\cos(\Omega-\theta)\csc i
    \end{array}
    \right] 
  \left[
    \begin{array}{cc}
      d X/d x \\
      d Y/d x
    \end{array}
    \right]
\end{array}
\end{equation}
and Eqs.~\ref{dfdx}--\ref{last}, \ref{dxdp}--\ref{dyde}.  Now we
have introduced two additional nonlinear parameters, $\Omega$ and $i$.

\begin{equation}
\begin{array}{l}
  \left[
    \begin{array}{c}
      d S/d \Omega \\
      d T/d \Omega
    \end{array}
    \right] = \\ \\ \left[
    \begin{array}{cc}
      -\sin(\Omega-\theta)\csc i & -\cos(\Omega-\theta)\cot i \\
      -\cos(\Omega-\theta)\cot i & \phantom{-}\sin(\Omega-\theta)\csc i
    \end{array}
    \right] 
  \left[
    \begin{array}{cc}
      X\\
      Y
    \end{array}
    \right]
\end{array}
\end{equation}
and
\begin{equation}
\begin{array}{l}
  \left[
    \begin{array}{c}
      d S/d i\\
      d T/d i
    \end{array}
    \right] = \\ \\ \left[
    \begin{array}{cc}
      -\cos(\Omega-\theta)\csc i\cot i & \sin(\Omega-\theta)\csc^2 i \\
       \sin(\Omega-\theta)\csc^2 i & \cos(\Omega-\theta)\csc i\cot i
    \end{array}
    \right] 
  \left[
    \begin{array}{cc}
      X\\
      Y
    \end{array}
    \right]
\end{array}
\end{equation}

\section{Nonlinear terms}
\label{nonlinear}
\subsection{Sources of Nonlinearity}
There are several small nonlinear terms which are important at the
$m/s$ and $\mu$as level, especially for the nearby and high-proper
motion stars likely to be observed by SIM Lite. 

{\it Secular acceleration} --- A star with significant proper motion
will have the radial component of its space velocity change with
position on the sky, resulting in a secular change in the radial
velocity up to $\sim$1 m/s/yr for the most extreme cases.  Secular
acceleration is given, to first order, by:
\begin{equation}
\label{defrv}
  \dot{v}_r = D\mu^2
\end{equation}
where, $v_r$ is the bulk radial velocity of the
star,\footnote{This quantity $v_r$ defines the 
  true radial velocity of the system with respect to the Solar System
  barycenter. It differs from the spectroscopic parameter $\gamma$ in
  that the latter is often measured with respect to a fiducial
  frame and can include non-Doppler effects such as instrumental
  offsets, gravitational redshift, and convective blueshift.} $D$ is 
the star's distance\footnote{$D$ is distinguished here from the
  inverse parallax $\varpi^{-1}$ simply for convenience of units.} and
$\mu$ is the total proper motion in radians per unit time.  This term
will be absorbed into the linear parameter $d$, if present, and so
could be ignored.

{\it Parallax changes} --- The change in parallax of nearby stars due
to their radial velocity may be of order 0.3 $\mu$as/yr.  The
shape of the parallactic motion is also a function of position on the
sky, and thus of the proper motion.  These changes can be of order 5
$\mu$as/yr.  The radial velocity term is given by:
\begin{equation}
  \dot{\varpi}= -\varpi\frac{v_r}{D}
\end{equation}

{\it Proper motion changes} --- The flip side of secular acceleration
is proper motion change due to change in distance.
This effect can be of order 3 $\mu$as/yr and is given by:
\begin{equation}
  \dot{\vec{\mu}} = -\vec{\mu}\frac{v_r}{D}
\end{equation}

{\it Curvilinear effects} --- The curvilinear nature of spherical
coordinates can produce what are essentially nonlinear terms depending
on how astrometric displacements are defined.  These effects are on
the same order as the above proper motion changes, and more pronounced
near the poles.  For $\mu$as astrometry, it suffices to handle
these effects by employing a rectilinear grid:  

In this work, astrometric 
displacements labeled $\Delta \alpha \cos \delta$ and $\Delta \delta$
do not strictly refer to changes in the right ascension and
declination of the star, but refer to displacements along rectilinear
axes along those dimensions at the nominal position of the star.  That
is, if the unit vector pointing to the nominal position of the star
from the Solar System barycenter is defined: 
\begin{equation}
\label{defp}
\hat{p} = [\cos\alpha\cos\delta,\sin\alpha\cos\delta,\sin\delta]
\end{equation}
then the unit vectors pointing east and north are given by 
\begin{eqnarray}
\label{defalpha}
\hat{\alpha}&=&[0,0,1]\times\hat{p}\\
\label{defdelta}\hat{\delta}&=&\hat{p}\times\hat{\alpha}
\end{eqnarray}
and the astrometric displacements are given by
\begin{eqnarray}
\Delta\delta &\equiv& \hat{p}^\prime\cdot\hat{\delta}\\
\Delta\alpha \cos \delta &\equiv& \hat{p}^\prime\cdot\hat{\alpha}
\end{eqnarray}
where $\hat{p}^\prime$ is the displaced position of the star and
$\hat{\delta}$ and $\hat{\alpha}$ are constant.

{\it Interferomteric cross terms} --- The 1-D interferometric
measurement of astrometric displacement on the sky may be
complicated by the motion of the reference and target stars.  That is, the
calculation of $\theta$ in Eq.~\ref{defrho} may require proper motion advanced or parallax
corrected positions in a manner specific to the details of a
particular instrument's measurement of $\theta$.  These cross terms
are likely to be small, and so they can estimated and refined in the
same manner as the other nonlinear terms, if necessary.

{\it RV-astrometry cross terms} --- The RV semi-amplitude $K$ is
related to the astrometric semi-major axis, $a$, by the parallax,
and so $\varpi$ is not strictly a linear parameter in the combined
astrometry-RV problem.  This effect can be large if the parallax is
small or not known, and should not be ignored for any system. 

{\it Relativistic Terms} --- Gravitational deflection by Solar System
objects and relativistic stellar aberration produce large, time dependent astrometric
displacements that depend on the position of the star, and are
therefore slightly nonlinear.  Because these displacements can be calculated
to better than $\mu$as precision given an estimate of the star's
true position to arcsecond precision, these effects are ignored here.

\subsection{Incorporating Nonlinear Terms}
Implementing these small nonlinear effects in our model requires
good estimates of the astrometric parameters (when astrometric data is
available, these parameters can be estimated from a first-pass
solution assuming no planetary companions).  The system is then solved  
using these estimates to calculate the second-order terms above.
Below, we indicate these estimated astrometric parameters with a tilde
to distinguish them from the solved parameters.  These estimates can
then be iteratively refined if necessary, but convergence should be
very fast for SIM Lite data.

We can include these nonlinear terms by making the following
substitutions to the bottom 3 rows of the matrix \bF\ in
Eqs.~\ref{defastrof} \& \ref{bfrvastro2}:
\begin{eqnarray}
\label{deftilde}
  \tau_k-t_0 &\rightarrow& \left(1 - (\tau_k-t_0)\frac{v_r}{2\tilde{D}}\right)(\tau_k-t_0)\\
  \Pi_{\alpha,k} &\rightarrow& \left(1 -
  (\tau_k-t_0)\frac{v_r}{\tilde{D}}\right)\Pi_{\alpha,k}\\
  \Pi_{\delta,k} &\rightarrow& \left(1 -
  (\tau_k-t_0)\frac{v_r}{\tilde{D}}\right)\Pi_{\delta,k}
\end{eqnarray}
and in the definitions of $\Pi_k$ (Eqs.~\ref{defpi} \& \ref{defpi2}):
\begin{eqnarray}
  \Pi_{\alpha,k} &\rightarrow&\Pi_{\alpha,k} + (\tilde{\Delta
    \alpha_0}\cos\delta + (\tau_k-t_0) \tilde{\mu_\alpha})\vec{r}\cdot\hat{p}\\
  \Pi_{\delta,k} &\rightarrow&\Pi_{\delta,k} + (\tilde{\phantom{,\cos}\Delta
    \delta_0\phantom{e\delta}} + (\tau_k-t_0) \tilde{\mu_\delta})\vec{r}\cdot\hat{p}
\label{advancedalpha}
\end{eqnarray}
where the quantities in parentheses have units of radians.  The
estimated terms $\tilde{\Delta \alpha_0} \cos \delta$ and 
$\tilde{\Delta  \delta_0}$ will likely be zero at first, but may be
iteratively refined with the other estimated parameters.  

Finally, the secular acceleration can be accommodated by simply subtracting
off the appropriate, approximate linear trend to the RV data, or, if the
trend parameter is present, allowing it to be absorbed in $d$.  

This procedure of estimating and refining linear astrometric terms
will also work for the nonlinear term introduced by the appearance of
$\varpi$ in $H$ and $C$ in Eqs.~\ref{defHC} \& \ref{defHC}.
Alternatively, the parallax can be treated as a nonlinear
parameter from the outset and eliminated from the linear coefficients
in $\vecb$ entirely.  

\section{Implementation}
\label{improvement}
\subsection{Public code}
\label{code}
We have implemented these algorithms into a set of IDL software
routines which we have made available publicly for the fitting of
radial velocity data.\footnote{Available at http$:$//exoplanets.org/code/}  The software package includes \texttt{RVLIN},
the user-defined function that may be passed to \texttt{MPFIT}, and \texttt{RV\_FIT\_MP},
a ``wrapper'' routine that employs \texttt{MPFIT} to fit a multi-planet
Keplerian model to a user-supplied set of radial velocity data.  We
anticipate maintaining and improving this package, and eventually
incorporating astrometric and transit data analysis.  This code, or
components of it, are currently used by members of the California
Planet Search.  Below, we discuss our software package's performance. 

\subsection{Speed-up From the Use of Explicit Derivatives}
The use of explicit derivatives above in an LM code significantly speeds up the
algorithm by avoiding unnecessary calculation of numerical
derivatives.  To quantify this improvement, we tested two cases with
published RV data, the 2-planet system HD 217107 \citep{Vogt05} and
the 5-planet system 55 Cnc \citep{Fischer08} (in both cases assuming
no jitter).  We combined the Lick and Keck RV data sets for both
cases, solving for all $5n+2$ parameters (including the RV offset
between the two telescopes).  We ran over 2,500 trials, where in each
trial we started the search with different initial guesses for the
orbital elements, each randomly drawn from a normal distribution with
a width given by the uncertainty in each parameter and centered on its
best-fit value.\footnote{Although dynamical fits are more precise, here we
  are only concerned with the algorithm's convergence in the region of
  the global $\chi^2$ minimum for a purely Keplerian fit.  The initial
guesses for the parameters in this and the test in
\S~\ref{convergence} were thus drawn near their values at this (presumed)
global minimum.}  We compared the total time
taken for these trials on a 2.6 GHz Intel Core 2 Duo MacBook Pro
running IDL 7.0 using explicit derivatives to the time taken on the
same machine with same initial guesses using numerical
derivatives.

In the case of HD 217107, the use of explicit derivatives sped up the
calculation by a factor of 2.3.  The improvement in the case of 55 Cnc
was even larger, a factor of 4.  The total number of steps taken by
\texttt{MPFIT} to converge on a solution was similar in the cases with and
without explicit derivatives, indicating that our step sizes for the
numeric derivatives were well chosen.  

\subsection{Convergence Benefits of Exploiting Linear Parameters}
\label{convergence}
We employed a custom version of the multi-planet fitting routine
often employed by the California and Carnegie Planet search as a
baseline to test the improvement in convergence of the algorithm
described in this work.  This routine, which derives from code described
in \citet{Marcy92} and \citet{Valenti95}, is essentially an LM
algorithm for searching all $5n+1$ parameters with carefully chosen
step-sizes for numerical derivatives.\footnote{That is to say, the
  only significant differences between the baseline code and the code
  described in \S~\ref{code} is the exploitation of linear parameters.}

Using the same hardware described above, we again fit the
\citet{Fischer08} 55 Cnc RV data from two telescopes.  We drew the
initial guesses for every parameter from normal distributions centered
on the Keplerian best-fit values and with width given by $s\sigma_x$, where 
$\sigma_x$ is the parameter uncertainty quoted in \citet{Fischer08}
and $s$ is a scale factor.  We varied $s$ smoothly from 0 (where the
initial guesses were the best-fit values exactly) to 10 (where every parameter
is independently, randomly altered by $10\sigma_x$) over 4000 trials.  

We did not attempt to fit for the telescope offset at each trial, as
this particular routine was not optimized for such a task.  We also
did not vary the $\gamma$ parameter from its best-fit value (though we
did fit for it), as no uncertainties were given for $\gamma$ in
\citet{Fischer08}.   

We compared the results of this custom code to our LM code employing
linear parameters (\S\ref{code}) with both explicit and numerical
derivatives.  In some sense, this is not a fair test since the
linear parameter models do not require or accept initial guesses
for $K$ and $\omega$, and thus the initial guesses are closer to the
best-fit values (since they differ in only 15 dimensions space,
whereas the guesses for the nonlinear routine they differ in 25 dimensions.)
We thus also ran an additional test, where the fully-nonlinear routine was
provided the best-fit values for $K$ and $\omega$ for each planet, and
thus had only 15 parameters varied (but still had all 26 parameters to
fit).  With this exception, we provided the same initial guesses to
each of the four schemes in each trial.

We recorded the final $\chi^2$ reported by each routine for each trial and compared
this with $\chi^2_{\rm min}$, the best-fit value.  We deemed any trial
for which $(\chi^2-\chi^2_{\rm min}) < 2$ to be a ``successful''
convergence on the correct parameters.\footnote{$\chi^2_{\rm min}\sim
  2910$.  Our results are only weakly sensitive to the precise definition of
  ``successful''}  At each tested
value of $s$, we weighted the full set of trials by a Gaussian window 75 trials
wide (corresponding to $\sim 0.19 \sigma$) and calculated the fraction
on trials that were successful.

\begin{figure}
\begin{center}
\plotone{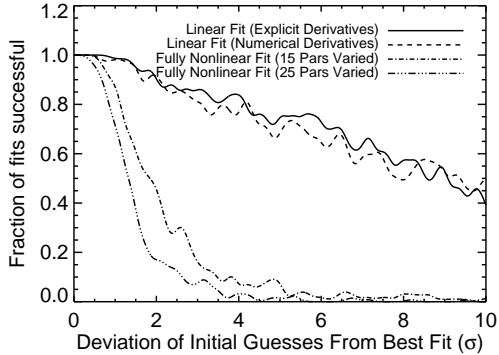}
  \caption{Sensitivity of RV-only fitting algorithms to initial guesses on
    the 5-planet, two-telescope RV data for 55 Cnc \citep{Fischer08}.
    Initial guesses to four fitting routines were randomly varied from their
    best-fit values by various factors of their respective
    uncertainties.  The linear parameter fitting algorithm
    described in this work converged on the best solution for a wide
    range of initial guesses, including $\sim 50\%$ of cases with guesses
    $10\sigma$ from nominal.  A full, 26-parameter nonlinear fit
    required guesses within 1.5-$\sigma$ of the best-fit value, or
    within $2\sigma$ if only the 15 truly nonlinear parameters were
    varied.}\label{stabletest} 
\end{center}
\end{figure}

In Figure~\ref{stabletest} we plot the fraction of successful tests as a
function of $s$.  The use of numerical derivatives had no overall effect on
the convergence of the linear pramater routine, but was slower,
on average, by a factor of 4.  

Even with initial guesses $10\sigma_x$ from the best-fit values in
all 15 parameters, the linear parameter routine found the global
minimum of $\chi^2$ in roughly half of all trials.  The
fully-nonlinear routine, searching a 26-dimensional space, 
required much better initial guesses to achieve convergence.  With the
same 15 parameters varied, fits with initial parameters off by
$3\sigma_x$  had only a 20\% chance of properly converging.  With all
25 orbital parameters varied, the initial
guesses had to be within 1.5-$\sigma_x$ of their proper value to have a
50\% chance of convergence.  With 26 dimensions to search, there are
many wrong paths for the LM algorithm to follow away from the global
$\chi^2$ minimum.

\section{Conclusions}
Applying linear parameters to the problem of fitting Keplerian curves
to radial velocity and astrometry data significantly improves the
efficiency and reliability of multi-planet Keplerian 
fitting routines.  This technique can be applied to many methods of searching
this complex, nonlinear, multi-parameter $\chi^2$ space, including the
Levenberg-Marquardt method, Markov chain Monte Carlo algorithms, and
brute force approaches.  Table~\ref{summary} summarizes the various schemes used in this work.

We have identified the nonlinear terms relevant for $\mu$as astrometry and m/s radial
velocity work (such as that by SIM Lite and its supporting RV data), and
shown how to incorporate these terms into a linear parameter
scheme.  In \S\ref{explicit} we have provided analytic forms for explicit derivatives
relevant to various applications of the Kepler problem.  

In the case of RV-only data, use of explicit derivatives can speed up
a fitting routine by a factor of 2--4, depending on the number of
planets being fit.  Use of linear parameters greatly improves the
convergence properties of a multi-planet fitting routine.  In the case
of an actual 5-planet fit, a linear parameter model requires
initial guesses within $10\sigma_x$ of their correct values in only
$15$ parameters to have a 50\% chance of convergence, while a full,
26-parameter search requires all 26 parameters to be specified within
1.5-$\sigma_x$.   

The principle improvement from use of linear parameters comes from
reducing the search space for an $n$-planet model.  In the case of RV
fitting, the reduction is from exploiting the linearity of $2n+1$ of
the $5n+1$ fitted parameters, leaving only $3n$ nonlinear parameters to be fit
algorithmically.  The problem of fitting astrometric orbits can be
similarly treated by exploiting the linearity of $4n+5$ of the $7n+5$
parameters, leaving, again, only $3n$ nonlinear parameters to be with
a nonlinear fitting routine.  When combining RV and astrometric data, as will be
the case for SIM Lite, only $2n+6$ of the $7n+6$ model parameters are
usefully linear under the scheme described here, leaving $5n$
nonlinear parameters. 

We have derived a general expression for the explicit model
derivatives for models employing linear parameters in nonlinear fits,
appropriate for application in the LM method.  This result is general
and can be applied to problems beyond Keplerian fitting -- indeed to
any model with both linear and nonlinear parameters.

\acknowledgments
\phantom{e}
Eric Agol inspired \S\ref{Example} by pointing out how
to exploit the linear parameters in the problem of RV fitting in the context of
\texttt{MPFIT}.  He has provided substantial guidance and ideas for this
and future versions of our code.  We thank him, Eric Ford, Matthew
Muterspaugh, Tom Loredoo, and Alessandro Sozzetti for careful readings of this
manuscript and their many substantive and constructive suggestions.

Our membership on a SIM Planet-Finding Astrometry Analysis Team (PI
Matthew Muterspaugh) motivated \S\S\ref{General} \& \ref{nonlinear} of this work, and we
are grateful to all of the Teams involved for their efforts and
insights into the problem.  We thank Jeff Valenti and Nikolai
Piskunov for encouraging us to calculate the explicit derivatives for a
general linear least-squares problem.  Nikolai Piskunov provided
insight into the sources of instability in the LM search algorithm.
Matthew Muterspaugh provided guidance with the problem of fitting
astrometric orbits and understanding the small astrometric cross
terms.  We thank Martin Sirk, Sam Halverson, and John Asher Johnson
for their work employing and debugging various versions of our code.
J.T.W received support from NSF grant AST-0504874, and A.W.H. received
support from NASA contract NAS7-03001 (JPL\#1336910).

\newpage
\begin{deluxetable}{lccc}
\tabletypesize{\scriptsize}
\tablecolumns{4}
\tablewidth{0pc} 
\tablecaption{Summary of Linearization Schemes\label{summary}}
\tablehead{ 
    \colhead{} & 
    \colhead{} &
    \colhead{} & 
    \colhead{Radial Velocities } \\
    \colhead{} & 
    \colhead{Radial Velocities} &
    \colhead{Astrometry} & 
    \colhead{\& Astrometry} 
}
\startdata 
Inputs &
    $\vecv(\vect),\tilde{\mu}, \tilde{\varpi},$ &
    $\vec{\Delta \alpha}(\vec{\tau}), \vec{\Delta  \delta}(\vec{\tau}), \tilde{\mu_\alpha}, \tilde{\mu_\delta}, v_r, \tilde{\varpi}$ &
    $\vecv(\vect), \vec{\rho}_{\vec{\theta}}(\vec{\tau}), \tilde{\mu_\alpha}, \tilde{\mu_\delta}, v_r, \tilde{\varpi}$ \\
Nonlinear parameters &
    $P_j, t_{{\rm p},j}, e_j$ &
    $P_j, t_{{\rm p},j}, e_j$ &
    $P_j, t_{{\rm p},j}, e_j, \Omega_j,i_j$ \\
Transformed linear parameters &
    $c_j, h_j, v_0 \rightarrow\omega_j, K_j, \gamma$ &
    $A_j, B_j, F_j, G_j \rightarrow \omega_j, \Omega_j, i_j, a_j$ &
    $C_j, H_j, v_0 \rightarrow \omega_j, \frac{m_j^3}{(m_*+m_j)^2}, \gamma$ \\
Linear parameters &
    $d$ &
    $\Delta\delta_0, \Delta\alpha_0, \mu_\delta, \mu_\alpha, \varpi$ &
    $d, \Delta\delta_0, \Delta\alpha_0, \mu_\delta, \mu_\alpha, \varpi$ 
\enddata
\tablecomments{The transformed linear parameters $v_0$ and $\gamma$ and all of the
  linear parameters appear once per system.  The subscript $j$ on the other parameters indicates that
  there are $n$ such parameters, one for each companion in the system.}
\end{deluxetable}

\begin{deluxetable}{clc}
\tablewidth{7in}
\tablecolumns{3}
\tablecaption{Variables and symbols used in this manuscript}
\tablehead{\colhead{Symbol} & \colhead{Meaning} & \colhead{Example equation}}
\startdata
\label{variables}
\scriptsize
$\tilde{\phantom{.}}$ & A tilde indicates a first approximation as
opposed to a fitted parameter &
\ref{deftilde} \\
$A,B,C,F,G,H$ & Thiele-Innes constants & \ref{defTI}\\
$a$ & Astrometric semi-major axis of a star's orbit in units of arc & \ref{defastar}\\
$\alpha$ & Nominal right ascension of a system at the epoch of
observations, $t_0$ & \\
$\hat{\alpha}$ & Constant unit east vector & \ref{defalpha}\\
$\vecb$ & Vector of linear parameters & \ref{defbeta}\\
$c$ & Linear parameter in RV fitting corresponding to a component of
a planet's RV signature& \ref{bdef} \\
$d$ & Linear parameter in RV fitting corresponding to an RV trend & \ref{bdef} \\
$D$ & Distance to a system & \ref{deftilde}\\
$\Delta \alpha \cos \delta, \Delta \delta$ & Measured astrometric
displacements in units of arc in the $\hat{\alpha}$ and $\hat{\delta}$ directions & \ref{defda} \& \ref{defdd}\\
$\delta$ & Nominal declination of a system at the epoch of
observations, $t_0$ & \\
$\hat{\delta}$ & Constant unit north vector & \ref{defdelta}\\
$\delta_{kl}$ & Kronecker delta & \ref{defKron}\\
$\gamma$ & Constant (and often instrument dependent) offset in a set of RV data &  \ref{modeleq}\\
$E$ & Eccentric anomaly of a planet (a function of time)  &
\ref{edef}\\
$e$ & Eccentricity of a planet & \ref{modeleq}\\
$\bF$ & Matrix defined such that the model $\vecu=\vecb\bF$  & \ref{deff}
\& \ref{defastrof} \\
$f$ & True anomaly of a planet (a function of time) &
\ref{edef}\\
$\theta$ & Position angle of astrometric displacement such that
$\theta=0$ refers to $\Delta \delta$ & \ref{defr}\\
$h$ & Linear parameter in RV fitting corresponding to a component of
a planet's RV signature & \ref{adef}\\
$i$ &  Inclination of a planet's orbit with respect to the sky & \ref{defTI}\\
$j$ &  Subscript indicating a quantity corresponds to the $j$th
planet & \ref{modeleq}\\
$K$ & RV semi-amplitude of a star's orbit due to planet &  \ref{modeleq}\\
$k$ & Subscript indicating a quantity corresponds to the $k$th observation & \ref{chi2}\\
$\lambda$ & A combination of nonlinear orbital parameters & \ref{defHC}\\
$M$ & Mean anomaly & \ref{Kepler}\\
$m_*$ & Mass of the primary component of a binary or planetary system  &
\ref{defastar} \\
$m$ & Mass of a smaller component of a binary or planetary system &  \ref{defastar}\\
$\mu, \mu_\alpha , \mu_\delta$ & Proper motion.  The total proper
motion is given by $\mu^2 =  \mu_\alpha^2
+ \mu_\delta^2$ & \ref{defastrof}\\
$n$ & Number of planets in a system & \ref{modeleq} \\
$N$ & Number of observations being fit & \ref{deff}\\
$\chi^2$ & The statistic & \ref{chi2}\\
$P$ & Period of a planet & \ref{Kepler}\\
$\hat{p}$ & Unit nominal position vector of a star in barycentric equatorial coordinates & \ref{defp}\\
$\Pi_{\alpha} , \Pi_{\delta}$ & Functions (of time and $\vec{r}$)
describing unit parallactic motion  & \ref{defpi} \& \ref{defpi2}\\
$\pi$ & The mathematical constant &  \\
$\varpi$ & Parallax of a system & \ref{defastar} \& \ref{defastrof}\\
$R_x(\omega), R_z(\omega)$ & The 3-D rotation matrix about the $x$- or $z$-axis. & \ref{defr}\\
$\vec{r}, r_x, r_y,r_z$ & Observatory position in barycentric equatorial
coordinates (a function of time) & \ref{defpi} \& \ref{defpi2}\\
$\rho_\theta$ & Measured astrometric displacement in the direction of position
angle $\theta$ & \ref{defrho}\\
$s$ & Scale of random deviation of initial guesses from nominal in units of $\sigma_x$ & \S\ref{convergence} \\
$S,T$ & Astrometric terms in the \bF\ matrix & \ref{defst}\\
$\sigma$ & Measurement uncertainty & \ref{chi2} \& \ref{defKron}\\
$\sigma_x$ & Uncertainty in an orbital parameter, $x$ & \ref{chi2} \& \ref{defKron}\\
$t$ & Time of a radial velocity observation & \ref{modeleq}\\
$t_0$ & Fiducial time at the epoch of the observations & \ref{deftilde}\\
$t_{\rm p}$ & Time of periastron passage of a planet & \ref{Kepler}\\
$\tau$ & Time of an astrometric observation & \ref{defastrou}, \ref{defastrof}\\
$u$ & Model values, such as velocities or astrometric displacements &
\ref{modeleq} \& \ref{modeleqmat}\\
$v$ & Measured radial velocities & \ref{chi2}\\
$v_0$ & Linear parameter corresponding to a constant offset in a set of RV data & \ref{cdef}\\
$v_r$ & Radial velocity of a system's barycenter with respect to Solar System barycenter & \ref{defrv}\\
$\bW$ & Diagonal matrix containing weights of measured data &
\ref{defKron}\\
$X, Y$ & Elliptical rectangular coordinates of a planet (functions of time) & \ref{defx} \&
\ref{defy}\\
$x$ & As a variable, can stand for any parameter, such as $P_j$, $t_{{\rm p},j}$, or $e_j$ & \ref{dudx}\\
$\Omega$ & Position angle of the ascending (approaching) node of a planet & \ref{defTI} \\
$\omega$ & Argument of periastron of a planet's orbit & \ref{modeleq} \\
$\omega_*$ & Argument of periastron of a star's orbit due to a planet.
 $\omega=\omega_*+\pi$& \ref{defTI} 
\enddata
\end{deluxetable}


\begin{thebibliography}{}

\bibitem[{Bakos} {et~al.}(2009){Bakos}, {Torres}, {P{\'a}l}, {Hartman},
  {Kov{\'a}cs}, {Noyes}, {Latham}, {Sasselov}, {Sip{\H o}cz}, {Esquerdo},
  {Fischer}, {Johnson}, {Marcy}, {Butler}, {Isaacson}, {Howard}, {Vogt},
  {Kov{\'a}cs}, {Fernandez}, {Mo{\'o}r}, {Stefanik}, {L{\'a}z{\'a}r}, {Papp},
  and {S{\'a}ri}]{Bakos09}
{Bakos}, G.~{\'A}., {Torres}, G., {P{\'a}l}, A., {Hartman}, J., {Kov{\'a}cs},
  G., {Noyes}, R.~W., {Latham}, D.~W., {Sasselov}, D.~D., {Sip{\H o}cz}, B.,
  {Esquerdo}, G.~A., {Fischer}, D.~A., {Johnson}, J.~A., {Marcy}, G.~W.,
  {Butler}, R.~P., {Isaacson}, H., {Howard}, A., {Vogt}, S., {Kov{\'a}cs}, G.,
  {Fernandez}, J., {Mo{\'o}r}, A., {Stefanik}, R.~P., {L{\'a}z{\'a}r}, J.,
  {Papp}, I., \& {S{\'a}ri}, P. 2009, arXiv:0901.0282v1

\bibitem[{Balan} \& {Lahav}(2008){Balan} and {Lahav}]{Balan08}
{Balan}, S.~T., \& {Lahav}, O. 2008, ArXiv 0805.3532, 805

\bibitem[{Binnendijk}(1960){Binnendijk}]{Binnendijk60}
{Binnendijk}, L. 1960.
\newblock {Properties of double stars; a survey of parallaxes and orbits.},
  Philadelphia, University of Pennsylvania Press [1960]

\bibitem[{Butler} {et~al.}(2006){Butler}, {Wright}, {Marcy}, {Fischer}, {Vogt},
  {Tinney}, {Jones}, {Carter}, {Johnson}, {McCarthy}, and {Penny}]{Butler06}
{Butler}, R.~P., {Wright}, J.~T., {Marcy}, G.~W., {Fischer}, D.~A., {Vogt},
  S.~S., {Tinney}, C.~G., {Jones}, H.~R.~A., {Carter}, B.~D., {Johnson}, J.~A.,
  {McCarthy}, C., \& {Penny}, A.~J. 2006, \apj, 646, 505--522

\bibitem[{Casertano} {et~al.}(2008){Casertano}, {Lattanzi}, {Sozzetti},
  {Spagna}, {Jancart}, {Morbidelli}, {Pannunzio}, {Pourbaix}, and
  {Queloz}]{Casertano08}
{Casertano}, S., {Lattanzi}, M.~G., {Sozzetti}, A., {Spagna}, A., {Jancart},
  S., {Morbidelli}, R., {Pannunzio}, R., {Pourbaix}, D., \& {Queloz}, D. 2008,
  \aap, 482, 699--729

\bibitem[{Catanzarite}, {Law}, \& {Shao}(2008){Catanzarite}, {Law}, and
  {Shao}]{Catanzarite08}
{Catanzarite}, J., {Law}, N., \& {Shao}, M. 2008, ArXiv e-prints, 807

\bibitem[{Catanzarite}, {Zhai}, \& {Shao}(2009){Catanzarite}, {Zhai}, and
  {Shao}]{Catanzarite09}
{Catanzarite}, J., {Zhai}, C., \& {Shao}, M. 2009, In American Astronomical
  Society Meeting Abstracts, volume 213 of {\em American Astronomical Society
  Meeting Abstracts\/}, pp. 456.03--+

\bibitem[{Cumming} {et~al.}(2003){Cumming}, {Marcy}, {Butler}, and
  {Vogt}]{Cumming03}
{Cumming}, A., {Marcy}, G.~W., {Butler}, R.~P., \& {Vogt}, S.~S. 2003, In ASP
  Conf. Ser. 294: Scientific Frontiers in Research on Extrasolar Planets, pp.
  27--30

\bibitem[{Driscoll} \& {Fischer}(2005){Driscoll} and {Fischer}]{Driscoll05}
{Driscoll}, P., \& {Fischer}, D. 2005, In Bulletin of the American Astronomical
  Society, volume~37 of {\em Bulletin of the American Astronomical Society\/},
  pp. 1269--+

\bibitem[{Eriksson} \& {Lindegren}(2007){Eriksson} and {Lindegren}]{Eriksson07}
{Eriksson}, U., \& {Lindegren}, L. 2007, \aap, 476, 1389--1400

\bibitem[{Fischer} {et~al.}(2008){Fischer}, {Marcy}, {Butler}, {Vogt},
  {Laughlin}, {Henry}, {Abouav}, {Peek}, {Wright}, {Johnson}, {McCarthy}, and
  {Isaacson}]{Fischer08}
{Fischer}, D.~A., {Marcy}, G.~W., {Butler}, R.~P., {Vogt}, S.~S., {Laughlin},
  G., {Henry}, G.~W., {Abouav}, D., {Peek}, K.~M.~G., {Wright}, J.~T.,
  {Johnson}, J.~A., {McCarthy}, C., \& {Isaacson}, H. 2008, \apj, 675, 790--801

\bibitem[{Ford}(2004){Ford}]{Ford04b}
{Ford}, E.~B. 2004, In The Search for Other Worlds, S.~S. {Holt} and
  D.~{Deming}, eds., volume 713 of {\em American Institute of Physics
  Conference Series\/}, pp. 27--30

\bibitem[{Ford}(2008){Ford}]{Ford08b}
{Ford}, E.~B. 2008, \aj, 135, 1008--1020

\bibitem[{Hastings}(1970){Hastings}]{Hastings70}
{Hastings}, W. 1970, Biometrika, 57

\bibitem[{Konacki} \& {Maciejewski}(1999){Konacki} and
  {Maciejewski}]{Konacki99}
{Konacki}, M., \& {Maciejewski}, A.~J. 1999, \apj, 518, 442--449

\bibitem[{Konacki}, {Maciejewski}, \& {Wolszczan}(2002){Konacki},
  {Maciejewski}, and {Wolszczan}]{Konacki02}
{Konacki}, M., {Maciejewski}, A.~J., \& {Wolszczan}, A. 2002, \apj, 567,
  566--578

\bibitem[Levenberg(1944)Levenberg]{Levenberg44}
Levenberg, K. 1944, Quart. Appl. Math., 2, 164--168

\bibitem[{Marcy} \& {Butler}(1992){Marcy} and {Butler}]{Marcy92}
{Marcy}, G.~W., \& {Butler}, R.~P. 1992, \pasp, 104, 270--277

\bibitem[{Markwardt}(2009){Markwardt}]{Markwardt09}
{Markwardt}, C.~B. 2009, ArXiv 0902.2850

\bibitem[Marquardt(1963)Marquardt]{Marquardt63}
Marquardt, D. 1963, SIAM J. Appl. Math., 11, 431--441

\bibitem[Metropolis {et~al.}(1953)Metropolis, Rosenbluth, Rosenbluth, Teller,
  and Teller]{Metropolis53}
Metropolis, N., Rosenbluth, A.~W., Rosenbluth, M.~N., Teller, A.~H., \& Teller,
  E. 1953, Journal of Chemical Physics, 21, 1087--1092

\bibitem[{Press} {et~al.}(1992){Press}, {Teukolsky}, {Vetterling}, and
  {Flannery}]{Press92}
{Press}, W.~H., {Teukolsky}, S.~A., {Vetterling}, W.~T., \& {Flannery}, B.~P.
  1992.
\newblock {Numerical recipes in FORTRAN. The art of scientific computing},
  Cambridge: University Press, |c1992, 2nd ed.

\bibitem[{Scargle}(1982){Scargle}]{Scargle82}
{Scargle}, J.~D. 1982, \apj, 263, 835--853

\bibitem[{Sozzetti}(2005){Sozzetti}]{Sozzetti05}
{Sozzetti}, A. 2005, \pasp, 117, 1021--1048

\bibitem[{Valenti}, {Butler}, \& {Marcy}(1995){Valenti}, {Butler}, and
  {Marcy}]{Valenti95}
{Valenti}, J.~A., {Butler}, R.~P., \& {Marcy}, G.~W. 1995, \pasp, 107, 966--+

\bibitem[{Vogt} {et~al.}(2005){Vogt}, {Butler}, {Marcy}, {Fischer}, {Henry},
  {Laughlin}, {Wright}, and {Johnson}]{Vogt05}
{Vogt}, S.~S., {Butler}, R.~P., {Marcy}, G.~W., {Fischer}, D.~A., {Henry},
  G.~W., {Laughlin}, G., {Wright}, J.~T., \& {Johnson}, J.~A. 2005, \apj, 632,
  638--658

\bibitem[{Wright} {et~al.}(2007){Wright}, {Marcy}, {Fischer}, {Butler}, {Vogt},
  {Tinney}, {Jones}, {Carter}, {Johnson}, {McCarthy}, and {Apps}]{Wright07}
{Wright}, J.~T., {Marcy}, G.~W., {Fischer}, D.~A., {Butler}, R.~P., {Vogt},
  S.~S., {Tinney}, C.~G., {Jones}, H.~R.~A., {Carter}, B.~D., {Johnson}, J.~A.,
  {McCarthy}, C., \& {Apps}, K. 2007, \apj, 657, 533--545

\bibitem[{Wright} {et~al.}(2009){Wright}, {Upadhyay}, {Marcy}, {Fischer},
  {Ford}, and {Johnson}]{Wright09}
{Wright}, J.~T., {Upadhyay}, S., {Marcy}, G.~W., {Fischer}, D.~A., {Ford},
  E.~B., \& {Johnson}, J.~A. 2009, \apj, 693, 1084--1099

\end{thebibliography}
\end{document}